\documentclass[%
 reprint,
 amsmath,amssymb,
 aps,
 prl,
nolongbibliography,
]{revtex4-2}

\usepackage{graphicx}% Include figure files
\usepackage{subcaption}% Enable subcaptions
\usepackage{dcolumn}% Align table columns on decimal point
\usepackage{bm}% bold math
\usepackage{color}
\newcommand{\od}[2]{\frac{d #1}{d #2}}

\newcommand{\pd}[2]{\frac{\partial #1}{\partial #2}}
\newcommand{\pds}[2]{\frac{\partial^2 #1}{\partial {#2}^2}}

\newcommand{\Rey}{{\rm Re}}
\newcommand{\Wi}{{\rm Wi}}

\newcommand{\Bx}{\boldsymbol{x}}

\newcommand{\Bh}{\boldsymbol{h}}

\begin{document}

\preprint{APS/123-QED}

\title{Universal Mean Velocity Profile in Polymeric Flows at Maximum Drag Reduction}% Force line breaks with \\

\author{F. Serafini}
\email{francesco.serafini@uniroma1.it}
\author{F. Battista}%
\author{P. Gualtieri}%
\author{C.M. Casciola}%
\email{carlomassimo.casciola@uniroma1.it}
\affiliation{Department of Mechanical and Aerospace Engineering, Sapienza University of Rome}


%\affiliation{
%\affiliation{
%\affiliation{%



\begin{abstract}
%%%%
Provided a sufficient concentration of long-chain polymers in a Newtonian solvent, turbulent wall-bounded flows exhibit a universal state known as Maximum Drag Reduction (MDR). Through direct numerical simulations, we show that the wall-normal kinetic energy flux characterising Newtonian wall-bounded turbulence is suppressed at MDR, and that the polymers mainly sustain velocity fluctuations. In agreement with the experimental data, we derive a universal expression for the mean velocity profile no longer dependent on characteristic quantities of wall-bounded turbulence.
%%%%
\end{abstract}

\maketitle
%-------------------------------------------------------------------------
%-------------------------------------------------------------------
Great interest in the dynamics of polymer solutions comes from the recognized capability of the polymers to significantly alter the dynamics of the Newtonian solvent in which they are dissolved \cite{benzi2018polymers}. High molecular weight chains can either enhance mixing at very small Reynolds numbers by triggering the Elastic turbulence \cite{groisman2000elastic} but also drastically reduce momentum mixing in high Reynolds number wall-bounded flows, leading to a significant drag reduction \cite{graham2004drag,procaccia2008colloquium}. This peculiar behavior has attracted many scientists since the experimental evidence of polymer drag reduction in the $40$'s \cite{toms1977early}. The effect of polymers is associated with the mechanical interaction between the solvent and the polymer chains elongated by the velocity field \cite{steinberg2021elastic,choi2002turbulent,serafini2022drag,serafini2025role}. 
A widely documented feature of wall-bounded drag-reducing polymer flows is that the mean velocity profile lies within two universal states: the Prandtl-Karman law of Newtonian turbulent flow and the Maximum Drag Reduction (MDR) state \cite{virk1975drag}. Interestingly, the profile at MDR is also reported to follow a log law
\begin{equation}
	U^+ = 11.7 \ln y^+ - 17 \, ,
	\label{eq:virk}
\end{equation}
where $U^+=U/u_\tau$ is the normalized mean streamwise velocity and $y^+=y/l_*$ is the normalized wall-normal distance. The $+$ superscript denotes normalization with the friction velocity $u_\tau=\sqrt{\tau_w/\rho}$ ($\tau_w$ is the wall shear stress and $\rho$ the solvent density) and the wall unit $l_*=\nu/u_\tau$ ($\nu$ is the kinematic viscosity).
This universal state has attracted considerable interest because of its seeming independence from the polymer/solvent parameters. and has been the subject of many theoretical, experimental, and numerical analyses.
\citet{procaccia2008colloquium} defined maximum drag reduction as a marginal state between a turbulent and a laminar regime of a wall-bounded flow in which the turbulent fluctuations have the role of ``properly" extending the polymers. \citet{benzi2005identification} proposed the MDR velocity profile as an edge solution of the Navier-Stokes equations (with an effective viscosity profile) beyond which no turbulent solutions exist. 
\citet{dubief2022elasto} suggest that MDR corresponds to the characteristic friction scaling of the Elasto-inertial turbulence, namely a chaotic state driven by polymer dynamics observed for the first time in the experiments by \citet{samanta2013elasto}.

An object of debate is whether the MDR velocity profile is logarithmic as Virk proposed. By numerical simulations, \citet{white2018properties} argued that when the drag reduction is large, an inertially dominated logarithmic scaling region ceases to exist because the polymers replace the role of the inertial mechanism. According to the Authors, maximum drag reduction is attained only after the inertial sublayer is eradicated, and they showed that the velocity profiles obtained by their simulations are no longer logarithmic. Despite the Authors considering a limited value of Reynolds number, it is apparent that the mean velocity profile at MDR systematically deviates from Virk's asymptote also in experiments, see e.g. \cite{warholic1999influence,ptasinski2001experiments}. Another open point is whether the Reynolds stress, namely the correlation between streamwise and wall-normal velocity fluctuation responsible for the drag increase in turbulent flows, is zero or not at MDR. The scenario captured by the experiments is unclear: \citet{warholic1999influence} measured zero Reynolds stress, whilst in \citet{ptasinski2001experiments}'s experiments the Reynolds stress is still non-zero. Nonetheless, the mean velocity profiles in both experiments have a similar shape and are close to Virk's asymptote, eq. \eqref{eq:virk}.

The MDR regime is further investigated in the present Letter, aided by direct 
numerical simulation of a turbulent pipe flow. It is found that the wall-normal kinetic energy fluxes are suppressed by the polymers when MDR occurs. This allows the derivation of a closed expression of the mean velocity profile that well describes the existing experimental data 
and explains the observed deviations from Virk's profile.

%-------------------------------------------------------------------
%-------------------------------------------------------------------
The dynamics of a dilute polymer solution can be conveniently described via the coupling of the Navier-Stokes equation for a Newtonian solvent with the Lagrangian evolution of $N$ polymer chains. This approach has been validated against experimental data and revealed to be quantitatively accurate \cite{serafini2025role}. Polymers are modeled as dumbbells, namely two massless beads at position 
$\Bx_{1,2}$, connected by a nonlinear spring, exchanging friction forces with the solvent. Despite being rather simple, the dumbbell model is sufficiently accurate for predicting polymer dynamics in turbulent flows \cite{serafini2024polymers,ching2024less} at an affordable computational cost.
The effect of N polymers with contour length $L$ and beads' friction coefficient $\gamma$ on the solvent is related to the polymer end-to-end vector $\Bh$ and can be quantified via the extra-stress tensor
\begin{equation}
	T_{ij} =  \frac{\beta}{2\Rey \Wi} \sum_{n=1}^{N} \frac{h_i^{(n)}h_j^{(n)}}{1-h_i^{(n)^2}} \delta(\Bx-\Bx_c^{(n)}) ,
	\label{eq:ps}
\end{equation} 
where the Weissenberg number $\Wi$ is the polymer relaxation to fluid time scale ratio and $\beta$ is the ratio between the polymer bulk viscosity $\mu_{p,b}=c_0\gamma L^2$ and the solvent viscosity $\mu_0$.
The derivation is reported in previous studies \cite{serafini2023role,serafini2025role}, where it has also been shown that, provided a sufficient number of polymers, the different polymer features (contour length $L$, concentration $c_0$, and friction coefficient $\gamma$) only affect the dynamics through the two dimensionless parameters $\beta$ and $\Wi$. Hence, MDR requires an explanation in terms of these two parameters. The saturation of the effect with the Weissenberg number is the easiest to explain. Indeed, the effect of the polymers on the solvent is only due to the $N_e$ polymers that are fully extended by the velocity field \cite{serafini2025role}. For the fully extended polymers, $h_i^2\simeq1$, the elastic force is proportional to the velocity difference sampled by the two beads along the direction specified by the end-to-end vector \cite{serafini2022drag}. It follows that
\begin{equation}
	\frac{1}{\Wi} \frac{1}{1-h_p^2} \simeq h_p({u}_{2_p}-{u}_{1_p}) \simeq h_p h_q \pd{u_p}{x_q} \bigg|_{\Bx_c} \, ,
\end{equation} 
and the polymer stress can be written as 
\begin{equation}
	T_{ij} =  \frac{\beta}{2\Rey} \sum_{n=1}^{N_e}   h_i^{(n)}h_j^{(n)}h_p^{(n)}h_q^{(n)} \pd{u_p}{x_q} \delta(\Bx-\Bx_c^{(n)})\, .
	\label{eq:vs}
\end{equation} 
The expression \eqref{eq:vs} shows that the polymer contribution to the total stress of the fluid is viscous, with a tensorial form given by the polymer end-to-end vector $\Bh$. The dependence on the Weissenberg number is embedded in the number of extended polymers $N_e$ and can be removed in the limit $\Wi\to\infty$.
Indeed, the number of extended polymers $N_e$ is a function of the Weissenberg number that approaches an asymptotic value if $\Wi\to\infty$ \cite{serafini2025role}.
The saturation with the viscosity ratio $\beta$ is here investigated by direct numerical simulation.

\begin{figure}
	\centering
	\includegraphics[width=0.4\textwidth]{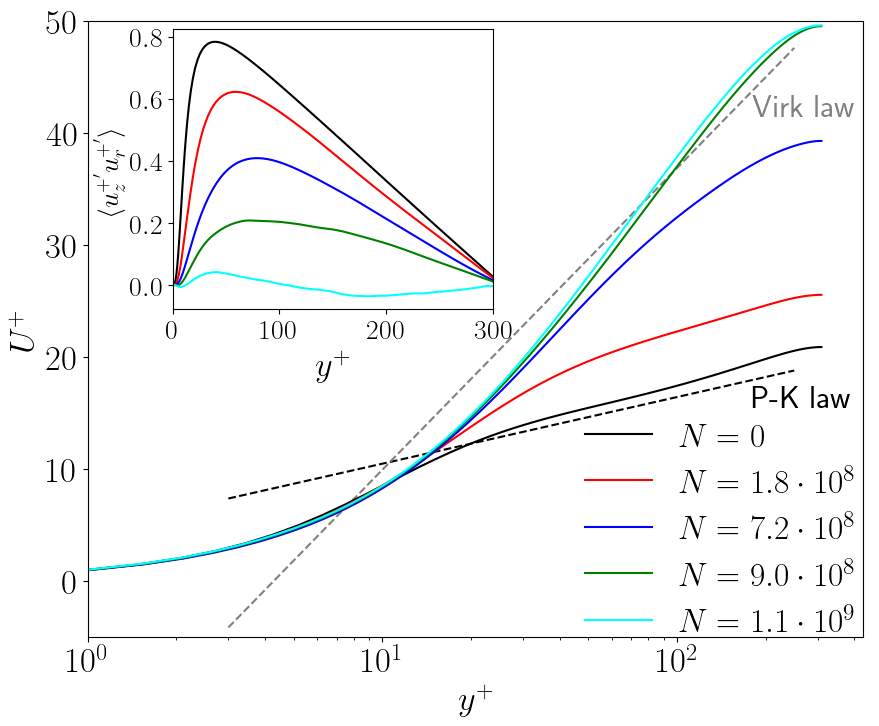}
	\caption{Mean velocity profiles and Reynolds stress profiles (inset) at constant friction Reynolds number $\Rey_\tau=310$ and Weissenberg number $\Wi=10^4$ for increasing number of polymers, see legend.
	}
	\label{fig:mvp}	
\end{figure}
Figure~\ref{fig:mvp} shows the mean velocity profile obtained by the numerical simulations with constant friction Reynolds number $\Rey_\tau=u_{\tau} R / \nu = 310$ ($R$ is the pipe radius), large Weissenberg number $\Wi=10^4$, and an increasing number of polymers to increase the concentration, i.e $\beta$. The mean velocity and the drag-reduction consequently grow with the concentration and approach an ultimate profile at the two largest values of concentration considered. Despite being close to it, the ultimate profile is different from Virk's asymptote. 

The inset of fig.~\ref{fig:mvp} reports the Reynolds stress $\langle u_z'^+ u_r'^+ \rangle$ measured in the different cases. Despite the velocity profiles at the two largest values of concentrations being equal within statistical accuracy, the Reynolds stress profiles are different: it is around zero at the largest concentration, whilst it is still noticeably larger in the other case. This anomaly is also reported in the experiments, see \cite{warholic1999influence,ptasinski2001experiments} as two examples. 
\begin{figure*}
	\centering
	\includegraphics[width=0.9\textwidth]{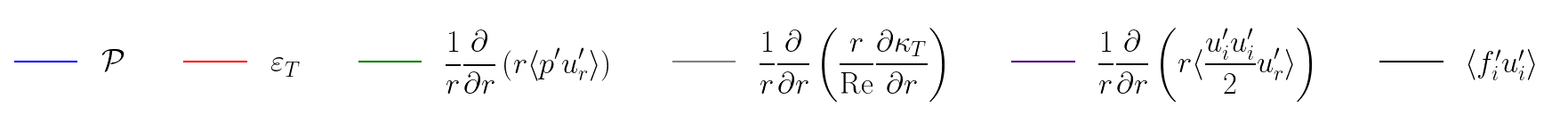}
	\includegraphics[width=0.4\textwidth]{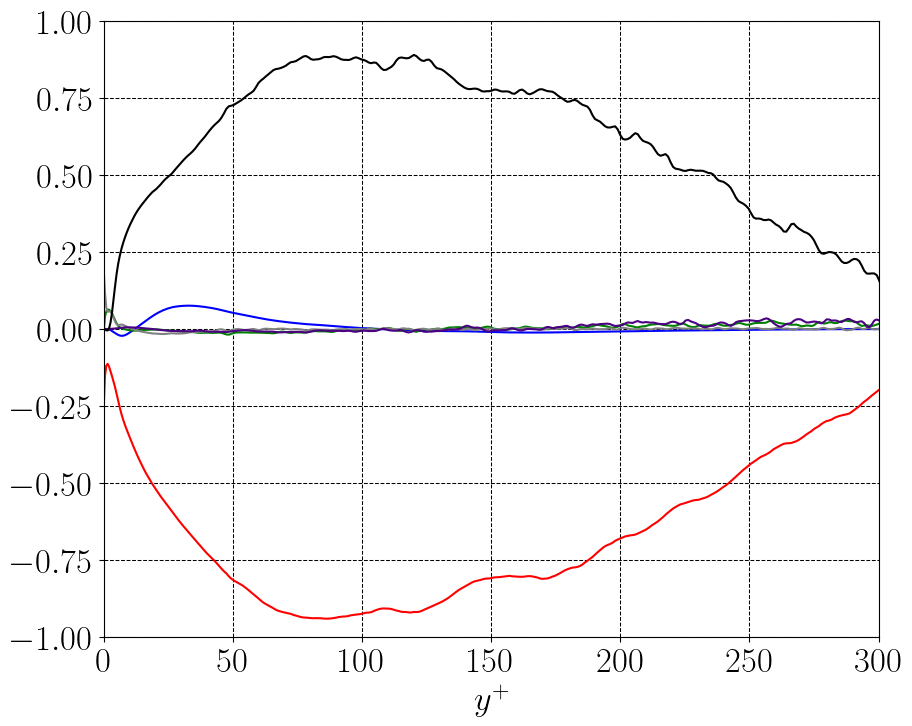}$
	{\scriptsize \put(-200,150){\bf (a)}}
	\includegraphics[width=0.39\textwidth]{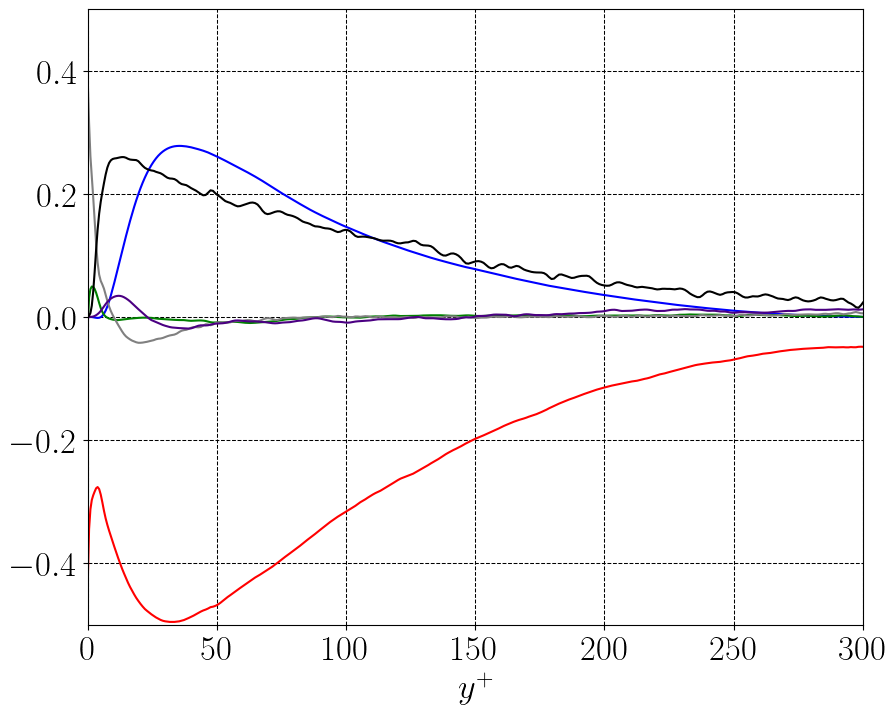}$
	{\scriptsize \put(-200,150){\bf (b)}}
	\caption{Turbulent kinetic energy budget for the two largest concentration simulations: $N=1.1\times 10^9$, panel (a), and $N=9.0\times 10^8$, panel (b).}
	\label{fig:tke}	
\end{figure*}
Consistently with the experimental measurements, the flow is turbulent as velocity fluctuations possess significant kinetic energy.  
The budget of turbulent kinetic energy $k_T=\langle u'_i u'_i \rangle /2$ is described by the equation
\begin{equation}
	\frac{1}{r}\pd{}{r} \left[ r \left( \langle p'u_r' \rangle + \langle \frac{u_i' u_i'}{2}u_r' \rangle -\frac{1}{Re} \pd{k_T}{r} \right) \right] = {\cal P} - \varepsilon_T + \langle f_i' u_i' \rangle \, ,
	\label{eq:ke}
\end{equation}
where ${\cal P}=-\langle u_z' u_r' \rangle \partial U/\partial r$ is the turbulent production, 
$\varepsilon_{T}= \langle \partial u_i'/\partial x_j \, \partial u_i'/\partial x_j \rangle /\Rey$ is the (pseudo)dissipation of fluctuating flow field, and $\langle f_i' u_i' \rangle$ is the power associated with the polymer forcing.
Figure~\ref{fig:tke} reports the turbulent kinetic energy budget for the two MDR simulations, the one with a larger polymer concentration in panel (a) and the one with a smaller polymer concentration in panel (b). Both panels show that the polymers are a source of turbulent kinetic energy, being $\langle f_i' u_i' \rangle>0$. It follows from the profiles of Reynolds stress in the inset of fig.~\ref{fig:mvp} that in the former case the turbulent production is negligible, whilst in the former case not. But aside from this difference, the two budgets show an important common feature: all the spatial flux terms are much smaller than the production/dissipation terms. On a more physical ground, the dissipation balances the production of turbulent kinetic energy, whatever kind it is, and no energy transfer occurs across the wall-normal direction. 
This observation suggests the definition of MDR as the state at which the polymer concentration is large enough to destroy the wall-normal turbulent transfer of kinetic energy, with most of the fluctuations induced by the polymers being locally dissipated.

\begin{figure}
	\centering
	\includegraphics[width=0.4\textwidth]{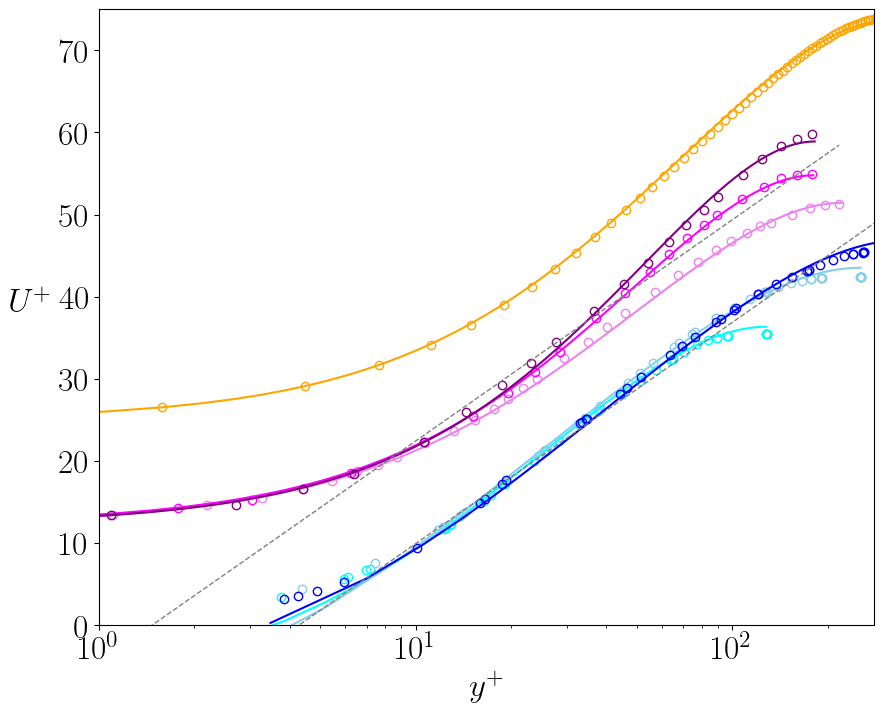}
	\caption{Comparison of the analytical mean velocity profile with present numerical data and experimental data. Numerical profile (orange circles) is compared with eq.~\eqref{eq:mdr} (orange line). Experimental data of \citet{ptasinski2001experiments} (blue tones circles) and of \citet{warholic1999influence} (violet tones circles) are compared with eq.~\eqref{eq:mdr4} (solid blue tone and violet lines, respectively). Dashed gray line is Virk's law \eqref{eq:virk}. To avoid the overlapping of the profiles, blue tone and orange data are shifted of $12.5$ and $25$ units, respectively.}
	\label{fig:mvp_c}	
\end{figure}

Even though a zero Reynolds stress appears not to be a requirement for MDR, in the case of negligible Reynolds stress, we can formulate a closed-form expression for the mean velocity profile. To do this, we express the polymer stress as a function of the mean velocity gradient via an effective viscosity $\mu_p(r)=\langle T_{rz} \rangle / (dU/dr)$. If the viscosity $\mu_p$ exhibits a simple dependence of the wall-normal distance $y=R-r$, e.g. linear as suggested in previous studies \cite{benzi2005identification}, an analytical form of the mean velocity profile can be written.
\begin{figure}
	\centering
	\includegraphics[width=0.4\textwidth]{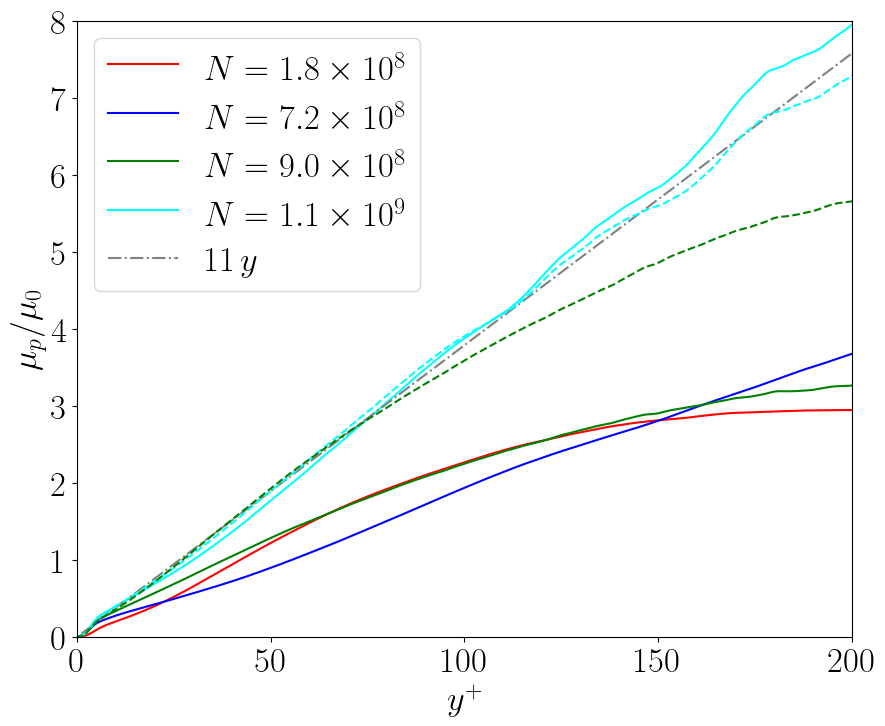}
	\caption{Ratio between mean polymer stress $\langle T_{rz} \rangle$ and mean velocity gradient (solid lines) and between the sum of polymer and Reynolds stress mean velocity gradient (dashed lines) for the different simulations, see legend.}
	\label{fig:visc}	
\end{figure}
We plot in fig.~\ref{fig:visc} the ratio between the polymer stress and the mean shear stress $\mu_0 dU/dy$ against the wall-normal distance in internal units $y^+$. This is the ratio between the polymer-induced viscosity profile $\mu_p(y)$ and the solvent viscosity $\mu_0$ and is linear with a good approximation for the MDR case with negligible Reynolds stress. In particular, $\mu_p/\mu_0 \simeq 11 \,y/R$.
Figure \ref{fig:visc} also plots the ratio between the sum of polymer and Reynolds stress and the viscous stress as dashed lines for the case $N=9\times 10^8$, the one with MDR velocity profile and non-zero Reynolds stress. Interestingly this quantity is linear and again comparable to $\mu_p/\mu_0 \simeq 11y/R$ for a wide range of wall-distances (up to $y^+\simeq100$). 
Using the hypotheses $\mu_p(y)/\mu_0 = \alpha y$ and $\langle u_z' u_r' \rangle = 0$, supported by numerical data, the mean momentum balance simplifies to 
\begin{equation}
	\frac{1}{r}\pd{}{r} \left\{ r\mu_0 \left[1+\alpha y \right] \pd{U}{r} \right\} = \od{p}{z}\bigg|_0 = 2 \frac{\tau_w}{R} \, .
	\label{eq:mmb}
\end{equation}
Double integration of~\eqref{eq:mmb} leads to the following mean velocity profile,
\begin{equation}
	U = \frac{\tau_w}{\mu_0\alpha} \left\{ \frac{1+\alpha R}{\alpha R} \ln (1+\alpha y) - \frac{y}{R} \right\} \, .
	\label{eq:mdr}
\end{equation}
The same expression is valid for the flow in a channel of half-height $R$. 
The expression \eqref{eq:mdr} perfectly matches the numerical data at MDR with the coefficient $\alpha=11$, see the orange line and symbols in fig.~\ref{fig:mvp_c}. Normalization with wall-units is still used to represent the data.
We also check the derived MDR profile against the experimental data. In particular, two experiments are considered: the one performed by \citet{warholic1999influence} (blue color tones) characterised by a zero Reynolds stress, and the one by \citet{ptasinski2001experiments} (violet color tones) who measured a non-zero Reynolds stress.
The profiles taken by \citet{ptasinski2001experiments} and present numerical data (orange tone) are intentionally shifted of $12.5$ and $25$ units in the ordinate axis for visualization purposes.
Experimental data are reported as empty circles, while the solid lines are the analytical profiles. The tricky point is extracting the coefficient $\alpha$ from the experimental data since the polymer stress is not directly available. The experimental data are rescaled using a viscosity $\mu_w$ that already accounts for the presence of the polymers, computed as the ratio between the wall-shear stress (measured via the pressured gradient) and the velocity gradient taken slightly away from the wall. Therefore, we introduce a viscosity profile $\mu_p(y)=\mu_w(1+\alpha\hat{y})$, with $\hat{y}=y-y_0$, and integrate the mean velocity profile from $y_0$ to $1$, 
\begin{equation}
	U =\frac{\tau_w}{\mu_w\alpha}\left\{ \frac{1+\alpha(1-y_0)}{\alpha} \ln (1+\alpha \hat{y} ) - \frac{\hat{y}}{R} \right\} + U(y_0) \, .
	\label{eq:mdr4}
\end{equation}
Using the value of $\mu_w$ given in the experiment, we can fit the value of $\alpha$ such that the fitted curves are those represented by solid lines (blue and violet tones) in fig.~\ref{fig:mvp_c}. 
The agreement with the experimental data of \citet{ptasinski2001experiments}, which are available at a smaller distance from the wall, is good in the entire range of normal wall distances, while the data of \citet{warholic1999influence}, with the first data available only at $4$ wall units, are well captured by the analytical curve for $y^+>10$. Overall, the accordance with the experimental measurements confirms the hypothesis of a mean linear viscosity brought by the polymers. 

With the derived expression validated against both numerical and experimental data, we now turn our attention to a more detailed examination of the velocity profile expression presented in eq.~\eqref{eq:mdr}. On a dimensional ground, the coefficient $\alpha$ is the inverse of a length scale, i.e. $\alpha=1/l_\alpha$. Hence, it is used to define a dimensionless wall-normal distance $y_\alpha = y/l_\alpha=\alpha y$. Looking at the derived analytical expression, we see that the mean velocity depends either on $y_\alpha$ or $\tilde{y}=y/R$. Finally, to obtain a dimensionless equation, we choose as reference velocity the quantity $U_0=l_\alpha \tau_w/\mu_0$, and we obtain
\begin{equation}
	\frac{U}{U_0} = U_\alpha(\tilde{y},y_\alpha) = \frac{1+R_\alpha}{R_\alpha} \ln (1+y_\alpha) - \tilde{y} \, ,
	\label{eq:mdr_4}
\end{equation}
with $R_\alpha=R/l_\alpha$.
For small $l_\alpha$, such that $R_\alpha \gg1$, there exists a range of wall distances for which $\tilde{y}\ll1$ and $y_\alpha\gg1$. Thus, if there is a large separation between the length scale $l_\alpha$ and the large scale $R$,  we recover a logarithmic law similar to the Prandtl-Karman and Virk log laws, $U_\alpha(y_\alpha) = \ln y_\alpha$. We observe that the choice of proper dimensional reference quantities suggested by the analytical expression allows us to derive a log law with no free coefficients.

To conclude, we show that, assuming as reference quantities the friction velocity $u_\tau$ and $l_*=\nu_0/u_\tau$, we recover the Virk's law \eqref{eq:virk} from the momentum balance eq.~\eqref{eq:mmb}. Virk's law is an experimental fitting and, as seen above, experimental data are typically rescaled using a viscosity taken at a distance $\delta$ from the wall. Assuming that the viscosity profile is linear, we have $\mu_\delta = \mu_0(1+\alpha\delta)$. Integration from $\delta$ the generic wall-normal distance $y$ leads to
\begin{equation}
	U = \frac{\tau_w}{\mu_0}\frac{1+\alpha\delta}{\alpha} \left\{ \frac{1+\alpha R}{\alpha R} \ln \frac{1+\alpha y}{1+\alpha \delta} - \frac{y}{R} \right\} + U(\delta) \, .
	\label{eq:mdrvd}
\end{equation}
Recalling the definitions of $u_\tau=\sqrt{\tau_w/\rho}$ and $l_*=\nu_0/u_\tau$ we obtain the following dimensionless expression of the mean velocity
\begin{equation}
	U^+ = \frac{1+\alpha^+\delta^+}{\alpha^+} \left\{ \frac{1+\alpha^+ \Rey_\tau}{\alpha^+ \Rey_\tau} \ln \frac{1+\alpha^+ y^+}{1+\alpha^+ \delta^+} - \tilde{y} \right\} + U^+(\delta^+) \, .
	\label{eq:mdrv}
\end{equation}
Since MDR requires $\alpha\gg1$, such that $\alpha^+\delta^+\gg1$ and $\alpha^+ y^+ \gg1$, in the limit of large Reynolds numbers there is large separation between $R$ and $l_*$, thus the mean velocity profile simplifies to 
\begin{equation}
	U^+ = \delta^+ \ln \frac{y^+}{\delta^+} + U^+(\delta^+) \, 
	\label{eq:mdr3}
\end{equation}
Choosing $\delta^+$ in the viscous sublayer (where $U^+=y^+$), then $U(\delta^+) = \delta^+$. Substitution into eq.~\eqref{eq:mdr3} leads to
\begin{equation}
	U^+ = \delta^+ \left( \ln y^+ - \ln \delta^+ + 1\right) =  \delta^+ \ln y^+ -\delta^+ \ln \frac{e}{\delta^+} \, , 
	\label{eq:mdrlog}
\end{equation}
which recovers the Virk's asymptote \eqref{eq:virk} for $\delta^+=11.7$.
We observe that the choice of $\delta^+$ relies on experimental data whose representation in $+$ units depends on the value of viscosity $\mu_w$ used to build the wall unit $l_*$. The analytical expression \eqref{eq:mdr_4} of the velocity profile instead clearly shows no dependence on the $+$ quantities characteristic of Newtonian wall-bounded turbulence. Normalization with wall units is not the natural choice for flows in the MDR regime, which are no longer related to the wall-bounded turbulence, as also proved by the depletion of the spatial fluxes of kinetic energy.

In summary, aided by direct numerical simulation data of drag-reducing polymer solutions in a pipe in the MDR regime, we observed that turbulent kinetic energy is no longer transferred across different wall-normal layers and that dissipation locally balances the production, mainly of polymeric origin.
The flow at MDR loses the typical structure of a wall-bounded turbulent flow. This allows the derivation of an analytical expression of the mean velocity profile if the additional hypothesis (supported by the data) of a linear viscosity profile induced by the polymers, $\mu_p/\mu_0=\alpha y$, is considered. The derived expression captures both numerical and experimental data.
The analytical results suggest that the proper normalization of the wall-normal distance and the mean velocity involves the length scale $l_\alpha=1/\alpha$ and the velocity scale $U_0=l_\alpha \tau_w/\mu_0$ and not the friction length and the friction velocity characterising wall-bounded Newtonian turbulence. The correct choice leads to the universal velocity profile in the MDR state, namely $U_\alpha=\ln y_\alpha$. 
%-------------------------------------------------------------------
%-------------------------------------------------------------------

%%%%%%%%%%%
%% Conclusions %%
%--------------------------------------------------------------------------

\appendix

\section{\label{sec:appA}Computational details}
The dimensionless system of equations describing the dynamics of the polymer solution is 
\begin{equation}
\begin{aligned}
	\pd{u_j}{x_j} = 0 \quad
	\pd{u_i}{t} +\pd{}{x_j} (u_i u_j) = -\pd{p}{x_i} + \frac{1}{\Rey} \pds{u_i}{x_j} + f_i\\
	\od{x_{c_i}}{t} = \frac{u_{1_i}+u_{2_i}}{2} + \frac{r_{eq}}{\sqrt{3\Wi}} \frac{\xi_{1_i}+\xi_{2_i}}{2} \\
	\od{h_i}{t} = \frac{u_{2_i}-u_{1_i}}{L} - \frac{h_i}{\Wi (1-h_i h_i)} 
	+ \frac{r_{eq}}{L\sqrt{3\Wi}} \left( \xi_{2_i} -\xi_{1_i}\right)
	\label{eq:NSfa}
\end{aligned} \, ,
\end{equation}
with the polymer forcing
\begin{equation}
	f_i =  \frac{\beta}{2\Rey \Wi} \sum_{n=1}^{N} \frac{h_i^{(n)}}{1-h_i^{(n)}h_i^{(n)}} \frac{\delta(\Bx-\Bx_1^{(n)}) -\delta(\Bx-\Bx_2^{(n)})}{L} \, .
	\label{eq:pf}
\end{equation}
In the equations \eqref{eq:NSfa} and \eqref{eq:pf}, $u$ is the fluid velocity, $p$ the hydrodynamic pressure, $x_c$ the position of the polymer centre, $h$ the polymer end-to-end vector (normalised by the polymer contour length $L$), $\xi$ a white noise setting the equilibrium size of the polymers $r_{eq}$, and $f$ the polymer backreaction on the solvent. The other dimensionless parameters appearing in the equations are the Reynolds number $\Rey$, the Weissenberg number $\Wi$, and the ratio of the polymer-induced viscosity to the solvent viscosity $\beta$.
The last two equations of system~\eqref{eq:NSfa} have to be integrated for all the $N$ dumbbells.

The dynamics of the polymer solution in a pipe is simulated by numerically solving the system~\eqref{eq:NSfa}, with the polymer forcing expressed by \eqref{eq:pf}, in cylindrical coordinates $(\theta,r,z)$ on a staggered grid with $ [N_\theta \times N_r \times N_z] = [2048 \times 329 \times 6192]$. 
Equations are solved in a cylindrical domain of size $(L_\theta,L_r,L_z)=(2\pi \times 1 \times 6\pi)$, with periodic boundary conditions in the streamwise directions and imposed pressure gradient to enforce a constant friction Reynolds number $\Rey_\tau=310$. Non-homogenous grid spacing in the radial direction guarantees a finer resolution at the wall, with a maximum grid spacing of $1$ wall units at the pipe centre.
The singular polymer forcing~\eqref{eq:pf} is regularised according to the Exact Regularised Point Particle method, see \cite{gualtieri2015exact,battista2019exact}.

\begin{acknowledgments}
Computational resources were provided by CINECA under the IscraB \#HP10BBQ4BV, EuroHPC grant EUHPC-R02-167, ICSC grant CNHPC-1954757.
This work has received financial support from Sapienza (Project RG12117A66DC803E) and ICSC – Centro Nazionale di Ricerca in “High-Performance Computing, Big Data and Quantum Computing”, funded by European Union – NextGenerationEU. 
\end{acknowledgments}

%%%%%%%%%%%%%%%%%%%%%%
%%% Bibliography 
%%%%%%%%%%%%%%%%%%%%%%
%\bibliography{my_library}

\end{document}